\documentclass[11pt]{article}
\usepackage{comment}
\usepackage[margin=2.5cm]{geometry}
\usepackage{amsmath,amssymb,extarrows,mathtools,graphicx,subfigure,setspace,bbold}
\usepackage{cite}
\usepackage{slashed}
\usepackage[dvipsnames]{xcolor}
\usepackage{hyperref}
\hypersetup{colorlinks=true, linkcolor=blue, citecolor=magenta, linktoc=page}
\usepackage{tikz}
\usepackage{arydshln,leftidx,mathtools}
\usetikzlibrary{decorations.pathreplacing}
\numberwithin{equation}{section}
\newcommand{\be}{\begin{equation}}
\newcommand{\bea}{\begin{eqnarray}}
\newcommand{\eea}{\end{eqnarray}}
\newcommand{\ba}{\begin{align}}
\newcommand{\ea}{\end{align}}
\newcommand{\ee}{\end{equation}}

\definecolor{c1}{rgb}{0.5,0,1}
\definecolor{c2}{rgb}{0.07,0.01,0.77}
\colorlet{aqua}{-red!75}

\begin{document}

\begin{titlepage}
\thispagestyle{empty}

\begin{flushright}
	IPM/P-2016/034
\end{flushright}

\vspace*{20mm}
\begin{center}
	{\Large {\bf Higher-curvature Corrections to Holographic Entanglement\\ with Momentum Dissipation}\\}
	
	\vspace*{15mm} \vspace*{1mm} {M. Reza Tanhayi${}^{a,b}$ and R. Vazirian${}^{a}$}
	
	\vspace*{1cm}
	
	${}^a\,\,${Department of Physics, Faculty of Basic Science, Islamic Azad
		University Central Tehran Branch (IAUCTB), P.O. Box 14676-86831,
		Tehran, Iran\\
		${}^{b}\,\,$School of Physics, Institute for Research in Fundamental Sciences
		(IPM) P.O. Box 19395-5531, Tehran, Iran}
	
	\vspace*{0.5cm}
	{E-mail: {\tt mtanhayi@ipm.ir}}
	
	\vspace*{1cm}
	%%\maketitle
\end{center}
\begin{abstract}
	
We study the effects of Gauss-Bonnet corrections on some nonlocal probes (entanglement entropy, $n$-partite information and Wilson loop) in the holographic model with momentum relaxation. Higher-curvature terms as well as scalar fields make in fact nontrivial correction to the coefficient of universal term in entanglement entropy.  We use holographic methods to study such corrections. Moreover,  holographic calculation indicates that mutual and tripartite information undergo a transition beyond which they identically change their values. We find that the behavior of transition curves depends on the sign of the Gauss-Bonnet coupling $\lambda$. The transition for $\lambda>0$ takes place in larger separation of subsystems than that of $\lambda<0$. Finally, we examine the behavior of modified part of the force between external point-like objects as a function of Gauss-Bonnet coupling and its sign.

\end{abstract}

\end{titlepage}

%\newpage

\tableofcontents
\noindent
\hrulefill

\onehalfspacing

%%%%%%%%%%%%%%%%%%%%%%%%%%%%%%%%
\section{Introduction}

The Anti-de Sitter (AdS)/Conformal Field Theory (CFT) correspondence postulates a relationship between quantum physics of strongly correlated many-body systems and the classical dynamics of gravity which lives in one higher dimension \cite{Aharony:1999ti}. Through this correspondence, a great deal of progress has been made in understanding the dynamics of strongly coupled gauge theories % On the other hand, in the reverse direction, this correspondence may improve our knowledge of geometry and quantum gravity \cite{Maldacena:2013xja}.
and it has also been further extended to cover topics related to the condensed matter theory \cite{Hartnoll:2011fn,Hartnoll:2009ns, Faulkner:2010da}. Actually, understanding the phenomena of
strongly coupled systems in condensed matter physics might be considered as one important goal of Gauge/Gravity duality.  Particularly, within the holographic point of view, much attention has
been paid to description of systems with momentum relaxation. Generically, in the gravity side, the solutions of Einstein-Maxwell-Dilaton theories have been frequently employed to address the states of underlying field theory. The solutions of such theories have in fact a net amount of charge and are fully translational invariant, so that, a small perturbation such as turning on an electric field, could result in an infinite DC
conductivity. It is obvious that such a model cannot present a realistic description of real physical systems.  In condensed matter materials  due to impurities or a lattice structure, the momentum is not conserved which this leads to a finite DC conductivity. Thus to give a realistic description
of materials in many condensed matter systems, translational symmetry must be broken. This can be done, for example, by breaking the translational invariance property \cite{Hartnoll:2007ih,Lucas:2014zea,Davison:2013jba,Horowitz:2012ky,Bao:2013fda,Mozaffar:2013bva,Lucas:2015vna,Andrade:2016tbr}. In this direction, Andrade and Withers presented a simple holographic model for momentum relaxation \cite{Andrade:2013gsa}. Their model consists of Einstein-Maxwell theory in ($d + 1$)-dimensional bulk space together with $d-1$ massless scalar fields. The neutral scalar fields in the bulk theory are dual to some operators with spatially dependent sources $\phi(x^i)$. These spatial sources can be chosen in a way that the bulk stress tensor and hence, the resulting black brane geometry are homogeneous and isotropic. Momentum relaxation concept is realized through these spatially dependent sources. Precisely, in this case it is shown that the holographic stress tensor obeys a conservation equation with contributions from the scalar vacuum expectation value $\langle O_\phi \rangle$
$$\nabla_i\langle T^{ij}\rangle=\langle O_\phi \rangle\nabla^j \phi, $$
which is indeed the modified Ward identity where $i, j$ label the boundary space-time directions. Noting that the Ward identity yields momentum conservation in the translationally invariant solution, its modification in this sense, results in breaking the translational invariance of the theory.\\
Inserting scalar fields into the theory in fact leads to deformation of  states at the corresponding dual field theory and it would be a relevant question that what happens to some specific concepts coming from holographic computation. For example some nonlocal measures of entanglement in such a model have been recently studied in \cite{Mozaffara:2016iwm} via the holographic methods.

The entanglement entropy in quantum field theories is an important quantity but difficult to compute in general. However, in strongly coupled field
theories, one can use holographic methods to calculate such nonlocal quantities. For example, to compute Holographic Entanglement Entropy (HEE) in the Einstein's theory of gravity, there is an elegant proposal made by Ryu and Takayanagi (RT) \cite{Nishioka:2009un}. According to the RT proposal, for a definite entangling region in the boundary, the entanglement entropy is related to the minimal surface ${\cal A}$, in the bulk whose boundary coincides with the boundary of the entangling region, 
\begin{equation}
S=\frac{{\cal A}}{4G_N},
\end{equation}
where $G_N$ stands for Newton's constant\footnote{In the extended version of RT proposal named as HRT proposal, for time-dependent geometries, one should use the extremal surface \cite{Hubeny:2007xt}.}. The above formula only works for CFT’s dual to Einstein gravity. In such theories, the central charges are same since in the gravity side, there are no extra parameters to distinguish the central charges. By expanding the parameter space of the couplings one can address this problem, which can be done by introducing higher derivative corrections in the action \cite{Henningson:1998gx, Abdalla:2001as,deBoer:2011wk,Ogawa:2011fw,Guo:2013aca,Alishahiha:2013dca}. Thus, to study general field theories in the context of holography, higher-derivative terms are in fact needed at the gravity side. In general, higher-derivative terms could potentially introduce ghost degrees of freedom; however, it is known that a special combination of curvature squared terms, namely the Einstein Gauss-Bonnet theory leads to second-order equations of motion and the theory is free of ghosts. Holographically, Gauss-Bonnet (GB) term plays the role of leading-order corrections to the Einstein gravity and in the context of AdS/CFT, GB background is dual to a theory with different central charges, \emph{i.e.}, $a$- and $c$-functions; it is noted that AdS solutions resulting from Einstein-Hilbert action yield the same $a$ and $c$ \cite{Myers:2012ed}. Motivated by the fact that adding higher-curvature terms into the action may help to investigate several new aspects of the theory, in this paper, we consider certain nonlocal probes of entanglement in momentum relaxation theories when the action contains GB term.  More precisely, we study the Holographic Entanglement Entropy (HEE), mutual and tripartite information; we also make a comment on potential between external objects by computing the expectation value of Wilson loop. We find the semianalytic expression for the coefficient of universal term in HEE which could introduce a modified `$c$'-type central charge in the corresponding dual quantum field theory.

In order to compute HEE in the semiclassical regime when some higher-order derivative terms are added into the Einstein gravity, RT proposal should be replaced by some other recipes  \cite{deBoer:2011wk, Hung:2011xb,Fursaev:2013fta,Dong:2013qoa,Camps:2013zua}. Some related works in this subject can also be found, for example in \cite{Mozaffar:2016hmg,Ghodsi:2015gna, Bueno:2014oua, Ling:2016dck} and references therein. 

In this paper, we will follow the proposal of \cite{Fursaev:2013fta} to study the HEE which will be reviewed in section 2. We will focus on GB gravity theory with momentum relaxation and compute the HEE for strip, spherical and cylindrical entangling regions in section 3. In section 4, other measurements of quantum entanglement in this setup will be considered, \emph{i.e.}, mutual and tripartite information and their quantum phase transitions and also the Wilson loop. In fact, we are interested in the effect of GB corrections to these quantities in holographic theories with momentum relaxation. The subject is concluded in section 5. Finally, in a short appendix we present some mathematical details.

%%%%%%%%%%%%%%%%%%%%%%%%%%%%%%%%
\section{Entanglement Entropy for Black Brane Solutions: A Short Review}

Entanglement entropy is an important nonlocal measure of different degrees of freedom in a quantum mechanical system \cite{Horodecki:2009zz}. This quantity similar to other nonlocal quantities, \textit{e.g.}, Wilson loop and correlation functions, can also be used to classify the various quantum phase transitions and critical points of a given system \cite{Vidal:2002rm}.\\
To define entanglement entropy in its spatial (or geometric) description, let us divide a constant time slice into two spatial regions $A$ and $B$ where they are complement to each other. Thus, the corresponding total Hilbert space can be written in a specific partitioning as
$\mathcal{H}=\mathcal{H}_A\otimes\mathcal{H}_{B}$. By integrating out the degrees of freedom that live in the complement of $A$, the reduced density matrix for region $A$ can be computed as $\rho_A={\rm Tr}_{B}\;\rho$ where $\rho$ is the total density matrix. The entanglement entropy is given by the Von-Neumann formula for this reduced density matrix as follows
\begin{equation}
S=-{\rm Tr}\;\rho_A \log \rho_A.
\end{equation}
For local $d$-dimensional quantum field theories, entanglement entropy follows the area law and it is infinite; the structure of the infinite terms are generally as follows \cite{Srednicki:1993im, Casini:2006hu, deBoer:2011wk}
\begin{align}
S(V)= \frac{g_{d-2}(\mathcal{A}_A)}{\epsilon^{d-2}}+\cdots+\frac{g_1(\mathcal{A}_A)}{\epsilon}+g_0(\mathcal{A}_A)\ln\epsilon+s(V),
\end{align}
where $\epsilon$ is the UV cutoff, $\mathcal{A}_A$ and $V$ stand for the area and volume of the entangling region in the boundary, $s(V)$ is the finite part of entropy and $g_i(\mathcal{A}_A) $ are local and extensive functions on the boundary of entangling region, which are homogeneous of degree $i$. The coefficient of the most divergent term is proportional to the area of the entangling surface and this is indeed the area law which is due to the infinite correlations between degrees of freedom near the boundary of entangling surface. The coefficients of infinite terms are not physical whereas the coefficient of logarithmic term is physical and universal in a sense that it is not affected by cutoff redefinitions.

Although computing the entanglement entropy in the context of field theory is indeed a difficult task, thanks to the AdS/CFT correspondence one can use RT proposal to find HEE. However, as mentioned in the introduction, for actions with higher-derivative terms, one should use other proposals to compute HEE. For example, in the case of curvature squared terms with the following action
\begin{equation}\label{action}
{\cal I} = \frac{1}{{16\pi {G_N}}}\int\limits_M {{d^{d + 1}}x} \sqrt { - g} \left[ {R - 2\Lambda  + a{R^2} + b{R_{\mu \nu }}{R^{\mu \nu }} + c{R_{\mu \nu \rho \sigma }}{R^{\mu \nu \rho \sigma }} - \frac{1}{2}\sum\limits_{i = 1}^{d - 1} {{{\left( {\partial {\phi _i}} \right)}^2}} } \right],
\end{equation}
pursuing the proposal of \cite{Fursaev:2013fta}, HEE is given by
\begin{multline}\label{fursaev}
S = \frac{{A\left( \Sigma  \right)}}{{4{G_N}}} + \frac{1}{4G_N} \int\limits_\Sigma  {\sqrt \sigma  {d^{d-1}x}} \left[ {2a R +b \left( {{R_{\mu \nu }}n_i^\mu n_i^\nu  - \frac{1}{2}\sum\limits_i {{{\left( {Tr{\mathcal{K}^{\left( i \right)}}} \right)}^2}} } \right)} \right. \\
\left. { + 2c \left( {{R_{\mu \nu \alpha \beta }}n_i^\mu n_i^\alpha n_j^\nu n_j^\beta  - \sum\limits_i {\mathcal{K}_{\mu \nu }^{\left( i \right)}\mathcal{K}_{\left( i \right)}^{\mu \nu }} } \right)} \right].
\end{multline}
In the above equations the cosmological constant is $\Lambda  =  - \frac{{d(d - 1)}}{{2{L^2}}}$,  ${{\phi _i}}$ are the minimally coupled massless scalar fields, $\sigma $ is the induced metric determinant, ${n_i}\,\left( {i = 1,2} \right)$ are the orthogonal normal vectors on the codimension two hypersurface $\Sigma $ and $\mathcal{K}_{\mu \nu }^{\left( i \right)}$ are the extrinsic curvature tensors on $\Sigma $ defined as
\begin{equation}\label{extrinsic-curvature}
\mathcal{K}_{\mu \nu }^{\left( i \right)} = h_\mu ^\lambda h_\nu ^\rho {\left( {{n_i}} \right)_{\lambda ;\rho }},\,\,\,\,\,\,\,\,\,\,\,\,\,\,\,h_\mu ^\lambda  = \delta _\mu ^\lambda  + \xi \sum\limits_i {{{\left( {{n_i}} \right)}_\mu }{{\left( {{n_i}} \right)}^\lambda }} ,
\end{equation}
where $\xi$ is $+1$ for time-like and $-1$ for space-like vectors. It is noted that the first term in \eqref{fursaev} is just the RT formula.\\
Corresponding equations of motion of \eqref{action} are given by
\begin{align}\label{eom-action}
&{\nabla _\alpha }{\nabla ^\alpha }{\phi _i} = 0, \notag\\
&{R_{\mu \nu }} - \frac{1}{2}{g_{\mu \nu }}R + \Lambda {g_{\mu \nu }} - \frac{1}{2}{g_{\mu \nu }}\left( {a{R^2} + b{R_{\alpha \beta }}{R^{\alpha \beta }} + c{R_{\alpha \beta \gamma \sigma }}{R^{\alpha \beta \gamma \sigma }}} \right) + 2a{R_{\mu \nu }}R - 4c{R_\mu }^\alpha {R_{\nu \alpha }} \notag\\
&+ \left( {2b + 4c} \right){R^{\alpha \beta }}{R_{\mu \alpha \nu \beta }} + 2c{R_\mu }^{\alpha \beta \gamma }{R_{\nu \alpha \beta \gamma }} + \left( {2a + \frac{b}{2}} \right){g_{\mu \nu }}{\nabla _\alpha }{\nabla ^\alpha }R + \left( {b + 4c} \right){\nabla _\alpha }{\nabla ^\alpha }{R_{\mu \nu }} \notag\\
&- \left( {2a + b + 2c} \right){\nabla _\nu }{\nabla _\mu }R + \sum\limits_{i = 1}^{d - 1} {\left( {\frac{1}{4}{g_{\mu \nu }}{\partial _\alpha }{\phi _i}{\partial ^\alpha }{\phi _i} - \frac{1}{2}{\partial _\mu }{\phi _i}{\partial _\nu }{\phi _i}} \right)}  = 0.
\end{align}
It is worth mentioning that the contribution of scalar fields to the stress tensor is supposed to be homogeneous, thus one gets a homogeneous and isotropic black brane solution. The solution can be written as
\begin{equation}\label{metric}
d{s^2} = \frac{{L^2}}{{{\rho ^2}}}\left( { - f\left( \rho  \right)d{t^2} + \frac{1}{{f\left( \rho  \right)}}d{\rho ^2} + \sum_{i=1}^{d-1} d{x_i}^2} \right),
\end{equation}
where $f(\rho)$ is a certain function of $\rho$ and we will return to this solution later. \\
Since we are specifically interested in studying the GB corrections in holographic theories with momentum relaxation, in what follows, we will limit ourselves to the five-dimensional GB gravity in the bulk with three specific scalar fields which are responsible for breaking the translational invariance in the dual field theory.
%%%%%%%%%%%%%%%%%%%%%%%%%%%%%%%%%%%%%%%%%%%%%%%%%%%%%%%%%%%%%%%%%%%%%
\section{Gauss-Bonnet Gravity with Linear Scalar Fields}

The GB gravity can indeed be obtained by setting $a = c =  - \frac{b}{4} \equiv \frac{\lambda  }{2}L^2$ in \eqref{action}, where $\lambda$ is a dimensionless coupling constant that controls the strength of
the GB term. The five-dimensional GB gravity is the simplest example of a Lovelock action and it is itself important because in a given background, the equations of motion for a propagating perturbation contain only two derivatives.\\ We work with the following Einstein GB scalar gravitational action
\begin{equation}\label{actionlam}
{\cal I} = \frac{1}{{16\pi {G_N}}}\int\limits_M {{d^5}x} \sqrt { - g} \left[ {R + \frac{{12}}{{L^2}} + \frac{{\lambda L^2}}{2}\left( {{R_{\mu \nu \rho \sigma }}{R^{\mu \nu \rho \sigma }} - 4{R_{\mu \nu }}{R^{\mu \nu }} + {R^2}} \right) - \frac{1}{2}\sum\limits_{i = 1}^3 {{{\left( {\partial {\phi _i}} \right)}^2}} } \right],
\end{equation}
where the action contains massless scalar fields to incorporate momentum relaxation in the system and they are considered to be linearly dependent on spatial coordinates, \emph{i.e.},
\begin{equation}\label{scalars}
{\phi _i} = {a_i}{x_1} + {b_i}{x_2} + {c_i}{x_3}.
\end{equation}
Such an ansatz for massless scalar sources, guarantees the solution to be homogeneous and isotropic. According to AdS/CFT dictionary, massless scalar fields are dual to marginal operators of the corresponding field theory and it was argued in \cite{Andrade:2013gsa} that such spatial dependent scalar field in the bulk modifies the Ward identity which leads to breaking of translational invariance in the dual field theory.

The relevant equations of motion for \eqref{actionlam} can simply be obtained from \eqref{eom-action} and the theory admits an asymptotically AdS$_5$ black brane solution as \eqref{metric} in which $f(\rho)$ is given by
\begin{equation}\label{frho}
f\left( \rho  \right) = \frac{{1 - \sqrt {1 - 4\lambda g\left( \rho  \right)} }}{{2\lambda }},
\end{equation}
where
\begin{equation}\label{g(rho)}
g\left( \rho  \right) = 1 - \frac{{{\alpha ^2}{\rho ^2}}}{4} - m{\rho ^4},\hspace*{1cm} m=\frac{1}{{{\rho _h}^4}}\left( {1 - \frac{{{\alpha ^2}{\rho _h}^2}}{4}} \right),
\end{equation}
with $\rho _h$ being the horizon radius and the constants $a_i,\,b_i$ and $c_i$ satisfy the following relations
\begin{align}\label{constraints}
&\sum\limits_{i = 1}^3a_i^2 =\sum\limits_{i = 1}^3  b_i^2 =\sum\limits_{i = 1}^3 c_i^2 = {\alpha ^2}, \nonumber\\
&\sum\limits_{i = 1}^3a_ib_i = \sum\limits_{i = 1}^3a_ic_i = \sum\limits_{i = 1}^3b_ic_i = 0.
\end{align}
It is noted that $f(\rho_h)=0$ and the UV boundary is defined as $\rho\rightarrow 0$ and the temperature of black brane is given by
\begin{equation}
T = \frac{1}{{\pi {\rho _h}}}\left( {1 - \frac{{{\alpha ^2}\rho _h^2}}{8}} \right).
\end{equation}
There is an interesting feature for the momentum relaxation methods, \emph{i.e.}, at the zero temperature one gets $f(\rho_h)=\frac{d}{d\rho}f(\rho)|_{\rho=\rho_h}=0$, which is an extremal black brane. Although there is no $U(1)$ charge to produce an extremal black brane solution in this case, the momentum relaxation parameter gives us such feature similar to the case of RN-AdS black brane.

%%%%%%%%%%%%%%%%%%%%%%%%%%%%%%%%%%%%%%%%%%%%%%%%%%%%%%%%%%
%\section{Holographic Entanglement Entropy in Gauss-Bonnet Gravity with Linear Scalar Fields}
%\subsection{Holographic Entanglement Entropy}
In the model that we are considering there are two deformations in the field theory due to the momentum relaxation parameter and GB term. In the following, we will develop the behavior of HEE of a quantum field theory whose states are in fact under the excitation of both momentum relaxation and GB term.

\subsection{HEE of a Strip}
In order to compute HEE, let us consider the following strip entangling region %\footnote{In this section we consider strip and generalization to the sphere and cylinder entangling regions would be straightforward. }
 \begin{equation}\label{strip}
 	- \frac{\ell}{2} < {x_1} \equiv x < \frac{\ell}{2},\,\,\,\,\,\,\,\,\,\,\,\,\,\,\, - \frac{{{H}}}{2} < {x_2}\,\,\text{and}\,\,{x_3} < \frac{{{H}}}{2},
 \end{equation}
where we assume $H\gg \ell$ and $H$ plays an infrared regulator distance along the entangling surface. The corresponding codimension two hypersurface in a constant time slice can be parametrized by ${x_1} = x\left( \rho \right)$; therefore, the induced metric becomes
\begin{equation}\label{induced}
 ds_{ind}^2 = \frac{{L^2}}{{{\rho ^2}}}\left[ {\left( {{{x'}^2} + {f^{ - 1}}} \right)d{\rho ^2} + d{x_2}^2 + d{x_3}^2} \right],
\end{equation}
where the $prime$ stands for the derivative with respect to $\rho.$ After doing some computation which are partially given in appendix, the entropy functional is found as follows\footnote{The GB gravity is a special form of curvature squared action and it was shown that for five-dimensional GB gravity, the proposal of computing HEE presented in \cite{Fursaev:2013fta} reduces to \cite{Hung:2011xb, deBoer:2011wk} and the results are the same. Note that taking into account the boundary term, only modifies the coefficient of leading UV-divergent term.}
\begin{equation}\label{efunc}
S = \frac{{{H^2}L^3}}{{4{G_N}}}\int d \rho \frac{{\sqrt {{{x'}^2} + {f^{ - 1}}} }}{{{\rho ^3}}}\left( {1 - 2\lambda \frac{{f\left( {fx'\left( {2\rho x'' + 3x'} \right) + 3} \right) - \rho f'}}{{{{\left( {1 + f{{x'}^2}} \right)}^2}}}} \right).
\end{equation}
The next step is minimizing the entropy functional \eqref{efunc} in order to find the profile of the hypersurface which has been parametrized by $x(\rho)$. It is noted that $x(\rho)$ is supposed to be a smooth differentiable function with the condition $x(0) = \ell /2$. To proceed, one may consider the entropy functional as a one-dimensional action in which the corresponding Lagrangian is independent of $x(\rho)$ which leads to a conservation law. In other words, let us write \eqref{efunc} as $S=\int d\rho {\cal L}$, thus the equation of motion becomes
\begin{equation}
\frac{\partial}{\partial \rho}(\frac{\partial {\cal L}}{\partial x''})-(\frac{\partial {\cal L}}{\partial x'})=C, \,\,\,\,\mbox{with}\,\,\,\,\frac{\partial {\cal L}}{\partial x}=0,
\end{equation}
where $C$ is a constant which can be fixed by imposing the condition that at the turning point $\rho_t$ of the hypersurface in the bulk one has $x'(\rho_t)\rightarrow \infty$.  After minimizing the functional of \eqref{efunc} and using the condition of the hypersurface turning point, one gets the following conserved quantity along the radial profile
\begin{equation}\label{eom}
x'\frac{{1+f\left( {{{x'}^2} - 2\lambda } \right)}}{{f{{\left( {{f^{ - 1}} + {{x'}^2}} \right)}^{3/2}}}} = \frac{{{\rho ^3}}}{{{\rho _t}^3}}.
\end{equation}
In principle, the above equation allows us to find $x'(\rho)$. In general, it is a difficult task to solve \eqref{eom} to find a proper profile since it is a cubic equation for $x'(\rho)$. However, in some special cases, the semianalytic solutions might be obtained. Up to the leading order of $\lambda$ and $\alpha$, and after making use of the following expression
\begin{equation}\label{ell}
\frac{\ell}{2}=\int_{0}^{\rho_t}x'(\rho)\,d\rho,
\end{equation}
one obtains
\begin{equation}
\ell=\frac{2 \sqrt{\pi } (1+\frac{3}{2}\lambda) \Gamma
	\left(\frac{2}{3}\right)}{\Gamma \left(\frac{1}{6}\right)}\rho_t+\frac{1}{12}
\alpha ^2 (1-\frac{3}{2}\lambda) \rho_t^3,
\end{equation}
which can be inverted to find the turning point of the proposed hypersurface in the bulk as follows
\begin{equation}
\rho_t=\frac{ \Gamma \left(\frac{1}{6}\right)}{2\sqrt{\pi } \Gamma
	\left(\frac{2}{3}\right)}\ell-\frac{ \Gamma
	\left(\frac{1}{6}\right)^4}{192 \pi ^2 \Gamma \left(\frac{2}{3}\right)^4}\ell^3\alpha ^2+\Big(\frac{ -3 \Gamma \left(\frac{1}{6}\right)}{4\sqrt{\pi } \Gamma
	\left(\frac{2}{3}\right)}\ell +\frac{ 5 \Gamma
	\left(\frac{1}{6}\right)^4}{128 \pi ^2 \Gamma
	\left(\frac{2}{3}\right)^4}\alpha ^2\ell^3\Big)\lambda +{\cal O}\left(\lambda ^2,\alpha^4\right).
	\end{equation}
Plugging the results into \eqref{efunc}, one gets the HEE as follows
\begin{equation}\label{EE}
 S = \frac{{{H^2}L^3}}{{4{G_N}}}\left(\frac{\mathfrak{a}}{{{\epsilon ^2}}} + \mathfrak{b}\log \frac{\ell }{\epsilon }+\frac{\mathfrak{c}}{\ell^2}\right)+{\cal O}(\lambda^2,\alpha^4),
\end{equation}
where $\epsilon$ stands for the UV scale which has been defined by the radial profile and
\begin{equation}
\begin{array}{l}
\mathfrak{a}=1-\frac{13}{2}\lambda,\\
\mathfrak{b}=\frac{1}{4}(1-\frac{3}{2}\lambda)\alpha^2,\\
\mathfrak{c}=-\frac{ 4\pi ^{3/2} \Gamma \left(\frac{2}{3}\right)^3}{ \Gamma
	\left(\frac{1}{6}\right)^3}(1+\frac{9}{2}\lambda)+\frac{\ell^2}{4}\Big((1-\frac{3}{2}\lambda)\log \frac{\Gamma
	\left(\frac{1}{6}\right)}{2^{2/3} \sqrt{\pi } \Gamma
	\left(\frac{2}{3}\right)}-\frac{1}{3}+ 3\lambda\Big)\alpha^2.
\end{array}
\end{equation}
The leading divergent term in \eqref{EE} is in fact the usual area law; on the other hand, the second term which is the universal logarithmic term is interesting. For a strip entangling region in CFT$_{d>2}$, in principle, there is no such a universal term in the HEE. Nevertheless, due to the momentum relaxation parameter and GB coupling, up to ${\cal O}(\lambda^2,\alpha^4)$ one obtains a logarithmic universal term as follows\footnote{This was first observed in \cite{Mozaffara:2016iwm} and \eqref{uni} is indeed its $\lambda$-correction. }
\begin{equation}\label{uni}
{S_{{\rm{univ}}{\rm{.}}}} = \frac{{{H^2}L^3}}{{16{G_N}}}(1 - \frac{3}{2}\lambda ){\alpha ^2}\log \frac{\ell }{\epsilon }.
\end{equation}
This term is physical and universal in a sense that it is not affected by cutoff redefinition and can be used to introduce a modified ‘c’-type central charge in the
corresponding dual quantum field theory. Using holographic entanglement entropy for strip geometry, Myers and Singh \cite{Myers:2012ed} introduced a candidate for a $c$-function in arbitrary dimensions. In a CFT$_4$ it goes as follows
 \begin{equation}\label{cfunction}
 c=\beta \frac{\ell^3}{H^2}\frac{\partial S(\ell)}{\partial \ell},
 \end{equation}
where $S(\ell)$ denotes the entanglement entropy for an interval of length $\ell$  and  the precise value of $\beta$ has been identified by holographic calculations which is given by
\begin{equation}
\beta=\frac{\Gamma
	\left(\frac{1}{6}\right)^3}{ 16\pi ^{3/2} \Gamma \left(\frac{2}{3}\right)^3}.
\end{equation}
In our setup by applying  \eqref{cfunction} and in the vicinity of $\lambda\backsimeq0$, one obtains
\begin{equation}\label{myers}
c=\frac{{ {L^3}}}{{8{G_N}}}\Big(1+\frac{9}{2}\lambda+\frac{\alpha ^2 \ell^2 (2-3 \lambda ) \Gamma
	\left(\frac{1}{6}\right)^3}{64 \pi ^{3/2} \Gamma
	\left(\frac{2}{3}\right)^3}\Big),
\end{equation}
 Note that by turning off the momentum relaxation one gets the modified version of this function due to the GB term at the linear level which was found in \cite{Myers:2012ed}.

%$$$$$$$$$$$$$$$$$$$$$$$$$$$$$$$$$$$$$$$$$$$$$$$$$$$$$$$$$$$$$$$$$$
\subsubsection{Low-Thermal Excitation ($m\ell^4\ll 1$) }

In a special case of setting $\alpha=0$, %\eqref{frho} reduces to
%\begin{equation}
%(\rho)=\frac{1}{2\lambda}\left(1-\sqrt{1-4\lambda(1-m\rho^4})\right),
%\end{equation}
we recover the five-dimensional GB AdS black brane solution with a Ricci-flat horizon which was found in \cite{Cai:2001dz}. On the other hand, for low-excited state of CFT and near the UV boundary, \eqref{metric} reduces to
\begin{equation}
ds^2=\frac{\tilde{L}^2}{\rho^2}\left(-g(\rho)d\tau^2+\frac{1}{g(\rho)}d\rho^2+dX_1^2+dX_2^2+dX_3^2\right),
\end{equation}
where $g(\rho)=1-m \rho^4+{\cal O}( m\lambda)$ and the modified AdS radius $\tilde{L}$ is defined by
\begin{equation}\label{ltil}
{{\tilde L}^2} = \frac{{L^2}}{{{f_\infty }}},\,\,\,\,\,\,{\rm{where}}\,\,\,\,{f_\infty } = \frac{{1 - \sqrt {1 - 4\lambda } }}{{2\lambda }}.
\end{equation} Excited state due to such deformation in CFT is called thermal excitation.  Thus, in particular, for the limit of $m\ell^4\ll 1$, the change of entropy can be obtained via the following relation
\begin{equation}\label{}
\Delta S=\frac{H^2\tilde{L}^2}{4G_N}\int d\rho(\frac{\delta\cal S}{\delta g} )|_{g=1}\Delta g,
\end{equation}
where ${\cal S}$, up to the leading order of GB coupling, is the integrand of entropy functional \eqref{efunc} given by
\begin{equation}
{\cal S}=\frac{{\sqrt {{{x'(\rho)}^2} + {g^{ - 1}(\rho)}} }}{{{\rho ^3}}}\left( {1 - 2\lambda \frac{{g(\rho)\Big( {g(\rho)x'(\rho)\left( {2\rho x''(\rho) + 3x'(\rho)} \right) + 3} \Big) - \rho g'(\rho)}}{{{{\left( {1 + g(\rho){{x'(\rho)}^2}} \right)}^2}}}} \right).
\end{equation}
Therefore, one obtains
\begin{equation}\label{deltas}
\Delta S = {S_{m \ne 0}} - {S_0} = \frac{{{H^2}{{\tilde L}^3}}}{{4{G_N}}}\frac{{(1 - 6\lambda {f_\infty })\Gamma {{\left( {\frac{1}{6}} \right)}^2}\Gamma \left( {\frac{1}{3}} \right)}}{{40\sqrt \pi  \Gamma {{\left( {\frac{2}{3}} \right)}^2}\Gamma \left( {\frac{5}{6}} \right)}}m{\ell ^2},
\end{equation}
where $S_0$ is the HEE for the vacuum case or pure AdS, namely $\alpha=\lambda=m=0$, and it is given by
\begin{eqnarray}\label{S0}
S_0=\frac{H^2}{4G_N}\Big(\frac{1}{ \epsilon ^2}-\frac{4}{\ell^2}\frac{ \pi ^{3/2} \Gamma \left(\frac{2}{3}\right)^3}{ \Gamma
	\left(\frac{1}{6}\right)^3}\Big).
\end{eqnarray}
The HEE \eqref{deltas} for low-thermal excitation due to the GB term will reproduce the result in \cite{Guo:2013aca}.

%%%%%%%%%%%%%%%%%%%%%%%%%%%%%%%%%%%%%%%%%%%%%%%%%%%%%%%%%%%%%%%%%%%%%%%%%%%%%%%%%%%%%%%%%%%%%%%%%%

\subsection{HEE of a Sphere}

In this case, let us use the metric \eqref{metric} with $\sum dx^2_i=dr^2+r^2d\Omega_2^2$ in which $f(\rho)$ is given by \eqref{frho}. On the boundary, the entangling region is a sphere with radius $r<\ell$, therefore, the corresponding codimension two hypersurface in the bulk is realized by $t=0$ and $r=F(\rho)$. With this assumption, the induced metric becomes
\begin{equation}\label{inducedsphere}
ds_{ind}^2 = \frac{{L^2}}{{{\rho ^2}}}\left[ {\left( {{{F'}^2} + {f^{ - 1}}} \right)d{\rho ^2} + r^2d{\theta}^2 + r^2\sin^2\theta d{\phi}^2} \right].
\end{equation}
Pursuing our previous example on strip entangling region, the corresponding entropy functional reads as
\begin{multline}\label{stripee}
S=\frac{\pi L^3}{G_N}\int d\rho \frac{F^2}{\rho ^3} \sqrt{\frac{1+f F'^2}{f}}\Bigg[1+\frac{2 \lambda }{\left(1+f F'^2\right)^2}\Bigg(f^2 F' \left(4 \rho  F'^2-2  \rho F  F''-3 F F'\right)\\+\rho f' \left(F-\rho  F'\right)+\frac{\rho^2 }{F}+f \rho
\left(4 F'-2 \rho  F''+\frac{\rho  F'^2}{F}\right)-3 fF\Bigg)\Bigg].
\end{multline}
By extremizing the obtained entropy functional and after making use of proper boundary conditions and for small parameters, the perturbative profile is found as follows\footnote{In the case of spherical and cylindrical entangling regions, we will consider the terms up to ${\cal O}(\lambda^2,\alpha^4,\lambda\alpha^2,m)$. }
\begin{multline}
F(\rho)=
\sqrt{\rho _t^2-\rho ^2}\Bigg[1-\frac{\lambda }{2}+\alpha^2\frac{1}{24 (\rho _t^2-\rho ^2){}^{3/2}}\Bigg(12 \sqrt{\rho
	_t^2-\rho ^2} \rho _t^4 \log (\frac{\rho _t}{\rho })\\+\left(2
\rho _t^2-\rho ^2\right) \Big(\sqrt{\rho _t^2-\rho ^2} \left(\rho ^2+5
\rho _t^2\right)+6 \rho _t^3 \log(\frac{\rho }{\sqrt{\rho _t^2-\rho
		^2}+\rho _t})\Big)\Bigg)\Bigg].
\end{multline}
From the identity of $\ell=F(0)$ one can obtain the turning point in terms of the entangling region length. Thus, semianalytic computation results in the following HEE for a sphere
\begin{equation}
S=\frac{\pi L^3}{2G_N}\Bigg[\frac{(2-13
	\lambda) \ell^2}{2 \epsilon ^2}-\frac{1}{4} (2-13 \lambda)-\frac{\alpha^2  \ell^2}{3}+\frac{1}{4} \left(30 \lambda +\alpha^2  \ell^2-4\right)
\log\frac{2 \ell}{\epsilon }\Bigg].
\end{equation}

%$$$$$$$$$$$$$$$$$$$$$$$$$$$$$$$$$$$$$$$$$$$$$$$$$$$
\subsection{HEE of a Cylinder}

In the case of the cylinder, let us parameterize the metric \eqref{metric} as
\begin{equation}\label{metriccylinder} 
d{s^2} = \frac{{L^2}}{{{\rho ^2}}}\left(  - f\left( \rho  \right)d{t^2} + \frac{1}{{f\left( \rho  \right)}}d{\rho ^2} + dz^2+dr^2+r^2d\phi^2 \right). 
\end{equation}
The entangling surface is a cylinder $r = \ell$ on the $t = 0$ surface in this boundary geometry. We also introduce a regulator length $H$ for the $z$ direction which is along the cylinder length. %The rest of our notation is the same as in the previous subsection.
By taking the profile as $r=F(\rho)$, the entropy functional becomes
\begin{multline}\label{cylinderEE}
S=\frac{{\pi H{L^3}}}{{2{G_N}}}\int d\rho\frac{F}{\rho ^3} \sqrt{\frac{1+f F'^2}{f}}(1-\lambda{\cal F}_{cyl.}),\\
{\cal F}_{cyl.}=\frac{2 f^2 F' \left(F \left(2 \rho  F''+3 F'\right)-2 \rho
	\left(F'\right)^2\right)+\rho  f' \left(\rho  F'-2 F\right)+2 f \left(\rho
	\left(\rho  F''-2 F'\right)+3 F\right)}{F\left(1+f F'^2\right)^2}.
\end{multline}
Minimizing the above expression results in the following equation of motion
\begin{multline}
2 f^3 F'^2 \Big(4 \lambda  \rho  F'^2+F \left(-4
\lambda  \rho  F''-6 \lambda  F'+3 F'^3\right)\Big)+2 f^2
\Big[F'\Big(\rho  F' \left(F'^2+2 \lambda \right)\\+6 F
\left(F'^2-\lambda \right)\Big)-\rho  F'' \left(6 \lambda
\rho  F'+F \left(F'^2-2 \lambda \right)\right)\Big]+\rho
\left(f' \left(2 \lambda  \rho -F F'\right)+2\right)\\-f \Big[4 \rho
\left(F'^2 \left(\lambda  \rho  f'-1\right)+\lambda \right)+F
\Big(F' \left(\rho  f' \left(F'^2-6 \lambda \right)-6\right)+2
\rho  F''\Big)\Big]=0,
\end{multline}
which should be solved to obtain a proper profile, which is indeed a difficult task. However, to identify the universal contribution, the near boundary behavior of the minimal surface would be sufficient. Thus, the asymptotic solution of this equation can be considered as
\begin{equation}
F(\rho)=c_1+c_2\rho+c_3\rho^2+\cdots,
\end{equation}
for which one finds
\begin{equation}
c_1=\ell,\,\,\,\,\,\,c_2=0,\,\,\,\,\,\,c_3=-\frac{1}{4\ell}+ \frac{\lambda}{4\ell}+{\cal O}(\lambda^2,\alpha^4),
\end{equation}
after making use of the boundary condition $F(0)=\ell$. By substituting the asymptotic form of the profile in \eqref{cylinderEE}, the universal part of HEE for the cylinder is finally obtained as follows
\begin{equation}
S_{{\rm{univ}}{\rm{.}}}=\frac{{\pi H{L^3}}}{{2{G_N}}}\,\,\frac{-2+7 \lambda+2\alpha^2  \ell^2 }{16 \ell}\log\frac{\ell}{\epsilon}.
\end{equation}

It is known that the holographic computation of Weyl anomaly could relate the gravity parameters $G_N$ and $\lambda$  to the central charges of dual CFTs \cite{Henningson:1998gx}. In two dimensions, central charge is related to the conformal anomaly via $\langle T^\mu_\mu \rangle= \frac{c}{24\pi} R$, where $R$ is the Ricci scalar. In principle there are two trace anomaly coefficients in 4-dimensional CFTs, namely $c$ and $a$-functions
\begin{equation}
\langle T^\mu_\mu \rangle\sim-a E_4+cW_{\alpha\beta\gamma\eta}W^{\alpha\beta\gamma\eta},
\end{equation}
where $E_4$ and $W_{\alpha\beta\gamma\eta}$ are respectively the Euler density  and Weyl tensor. In this way it is argued that for spherical entangling region the universal part of the HEE would be proportional to $a$  while for cylindrical entangling region it relates to $c$-function \cite{Solodukhin:2008dh}. However, for all AdS backgrounds, one obtains
\begin{equation}
\begin{array}{l}
S_{{\rm{univ}}{\rm{.}}}=-\frac{\pi L^3}{2G_N}\log\frac{2\ell}{\epsilon}\equiv-4a\log\frac{2\ell}{\epsilon},\hspace{16mm}\mbox{for sphere},\\
S_{{\rm{univ}}{\rm{.}}}=-\frac{\pi L^3 }{16G_N}\frac{H}{\ell}\log\frac{\ell}{\epsilon}\equiv-\frac{c}{2}\frac{H}{\ell}\log\frac{\ell}{\epsilon},\hspace{12mm}\mbox{for cylinder},
\end{array}
\end{equation}
which means for all theories dual to Einstein gravity one gets the same central charges namely $a=c=\frac{\pi L^3}{8G_N}$;  On the other hand, for higher curvature
gravity theories these coefficients get modified due to stringy corrections (see \cite{Buchel:2009sk} for GB gravity). In our setup, holographic computation shows that the coefficient of universal logarithmic terms have been modified as follows
\begin{equation}\label{charge}
\begin{array}{l}
`c`=c\Big(1-\frac{7}{2}\lambda-\alpha^2\ell^2\Big),\\
`a`=a\Big(1-\frac{15}{2}\lambda-\frac{1}{4} \alpha^2\ell^2\Big).
\end{array}
\end{equation}

%. In the case of  Gauss-Bonnet gravity the holographic calculation results in \cite{Buchel:2009sk}
%\begin{equation}
%\begin{array}{l}
%c=\frac{\pi \tilde{L}^3}{8G_N}\sqrt{1-4\lambda},\\
%a=\frac{\pi \tilde{L}^3}{8G_N}(-2+3\sqrt{1-4\lambda}).
%\end{array}
%\end{equation}
This means that deforming the theory by adding higher-curvature terms and some specific scalar fields to break the momentum conservation, results in a change of the universal terms of the dual field theory. And the corresponding coefficients may be interpreted as the corrected (modified) central charges of the dual theory.

%%%%%%%%%%%%%%%%%%%%%%%%%%%%%%%%%%%%%%%%%%%%%%%%%%%%%%%%%%%%%%%%%%%%
%%%%%%%%%%%%%%%%%%%%%%%%%%%%%%%%%%%%%%%%%%%%%%%%%%%%%%%%%%%%%%%%%%%%%%%%%%
\section{Holographic $n$-partite Information and Wilson Loop}

In addition to entanglement entropy, the $n$-partite information and also Wilson loop are in fact useful quantities developed in the framework of gauge/gravity duality. In the case of two and three entangling regions, the $n$-partite information is equivalent to holographic mutual and tripartite information, respectively. These quantities indicate the amount of shared information, or more precisely the correlation, between the entangling regions \cite{Bernamonti:2012xv}. On the other hand, Wilson loop is in fact another nonlocal operator which can be used as an important probe for studying phase structures of gauge theories. Investigating the effect of higher-order terms and momentum dissipation on these quantities is the main task of this section.

%%%%%%%%%%%%%%%%%%%%%%%%%%%%%%%%%%%%%%%%%%%%%%%%%%%%%%%
\subsection{Holographic Mutual Information}

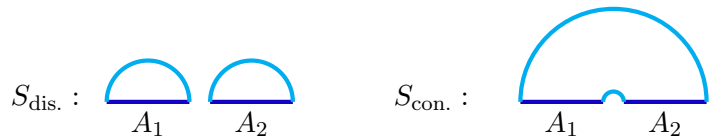
\begin{figure}[h!]
	\centering
	\begin{tikzpicture}[scale=.55]
	\draw[ultra thick,c2] (0,0) -- (2,0);
	\draw[ultra thick,c2] (2.5,0) -- (4.5,0);
	\draw[ultra thick,cyan] (2,0) arc (0:180:1cm);
	\draw[ultra thick,cyan] (4.5,0) arc (0:180:1cm);
	\draw[ultra thick,c2] (10,0) -- (12,0);
	\draw[ultra thick,c2] (12.5,0) -- (14.5,0);
	\draw[ultra thick,cyan] (14.5,0) arc (0:180:2.25cm);
	\draw[ultra thick,cyan] (12.5,0) arc (0:180:0.25cm);
	%\draw[step=1cm,gray,very thin] (0,-10) grid (25,25);
	\draw[] (-.4,0.1) node[left] {$S_{\text{dis.}}:$};
	\draw[] (9,0.1) node[left] {$S_{\text{con.}}:$};
	\draw[] (1,0) node[below] {$A_1$};
	\draw[] (3.5,0) node[below] {$A_2$};
	\draw[] (11,0) node[below] {$A_1$};
	\draw[] (13.5,0) node[below] {$A_2$};
	\end{tikzpicture}
	\caption{Schematic representation of two different configurations for computing $S\left( {{A_1} \cup {A_2}} \right)$. In the case of disconnected diagram: $ S({A_1} \cup {A_2})=2S(\ell)$ whereas for connected diagram: $ S({A_1} \cup {A_2})=S(2\ell+h)+S(h) $. }\label{figure1}
\end{figure}

\begin{figure}[h!]
	\centering
	\includegraphics[width=8cm]{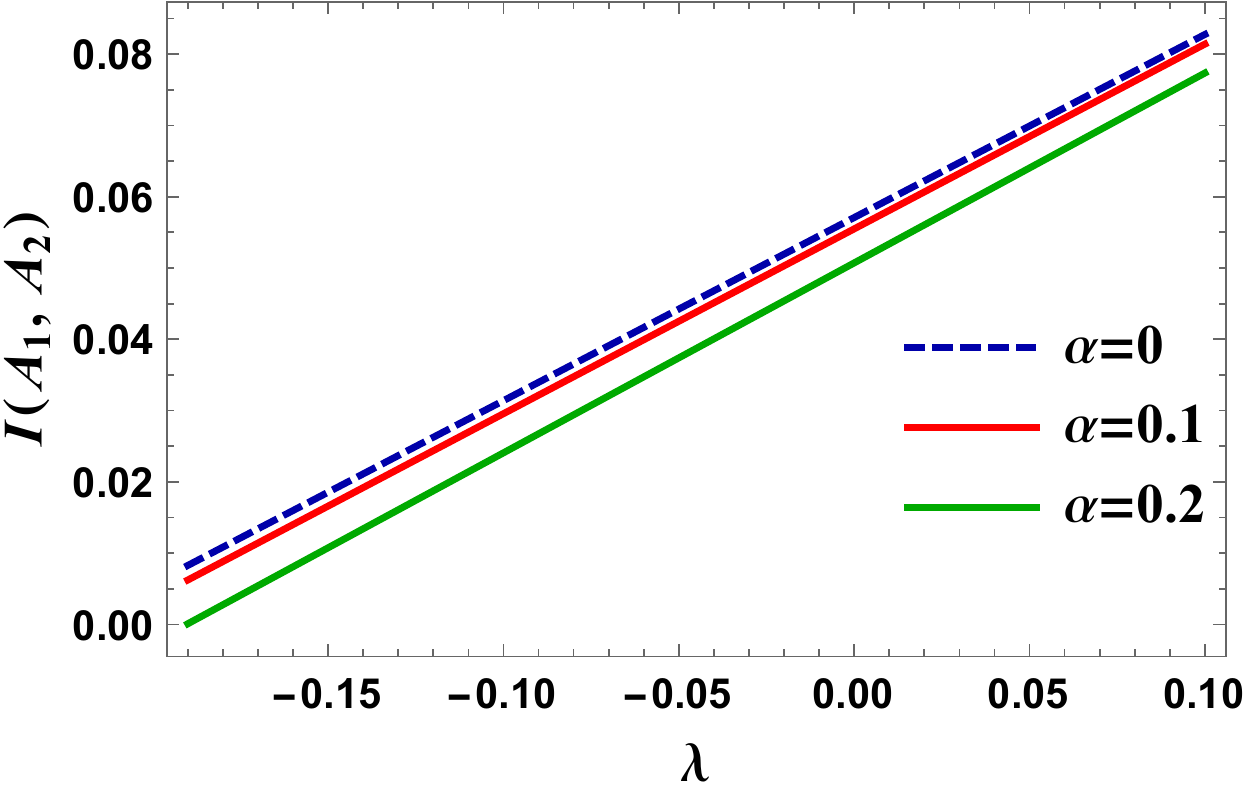}\,\,\,\,\,\,\,\,\,\,\,\,\includegraphics[width=8cm]{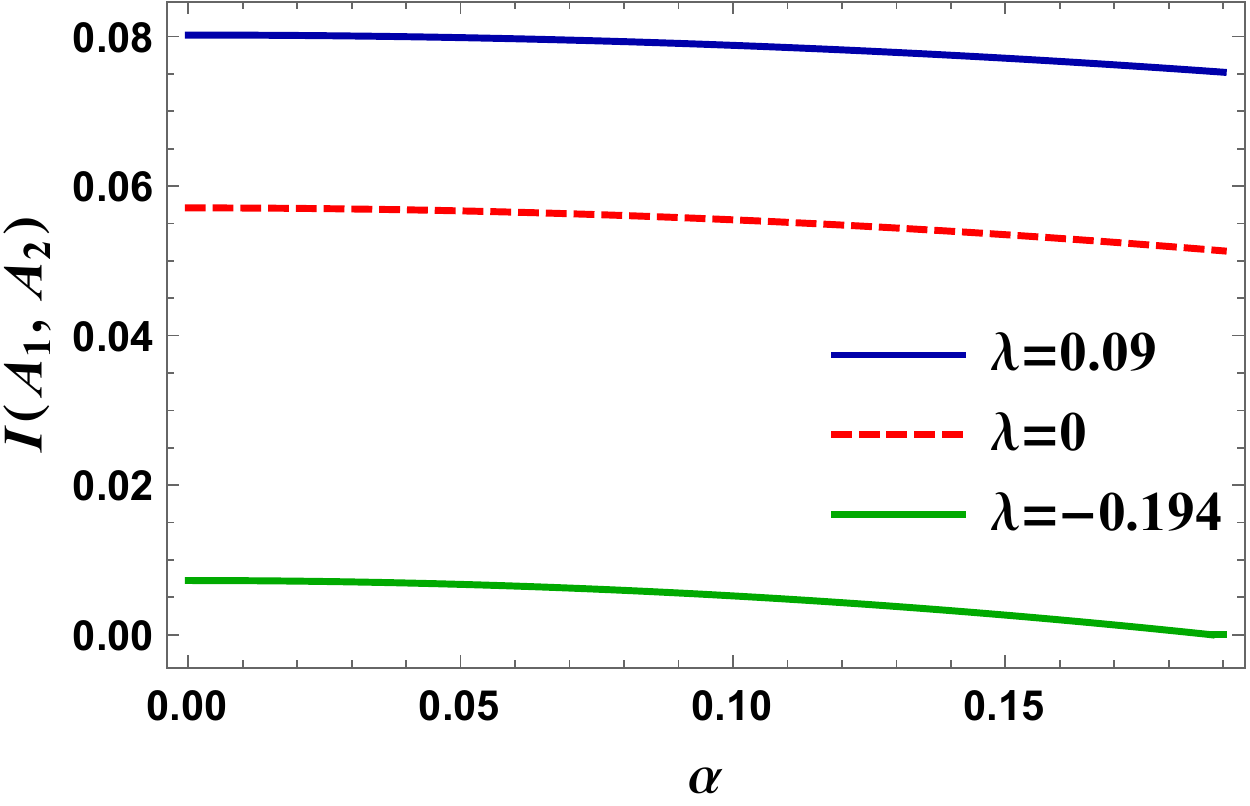}
	\caption{\textit{Left plot}: The behavior of mutual information as a function of GB parameter for different fixed values of $\alpha$. \textit{Right plot}: The behavior of mutual information as a function of momentum relaxation parameter for different fixed values of $\lambda$. It is noted that in both of the figures we set $\ell=1$ and $h=0.7$.  }%\textit{Right plot}: Transition of critical point as a function of $\lambda$ when $\alpha=0.1$. }
	\label{mutualfig}
\end{figure}

\begin{figure}[h!]
	\centering
	\includegraphics[width=8cm]{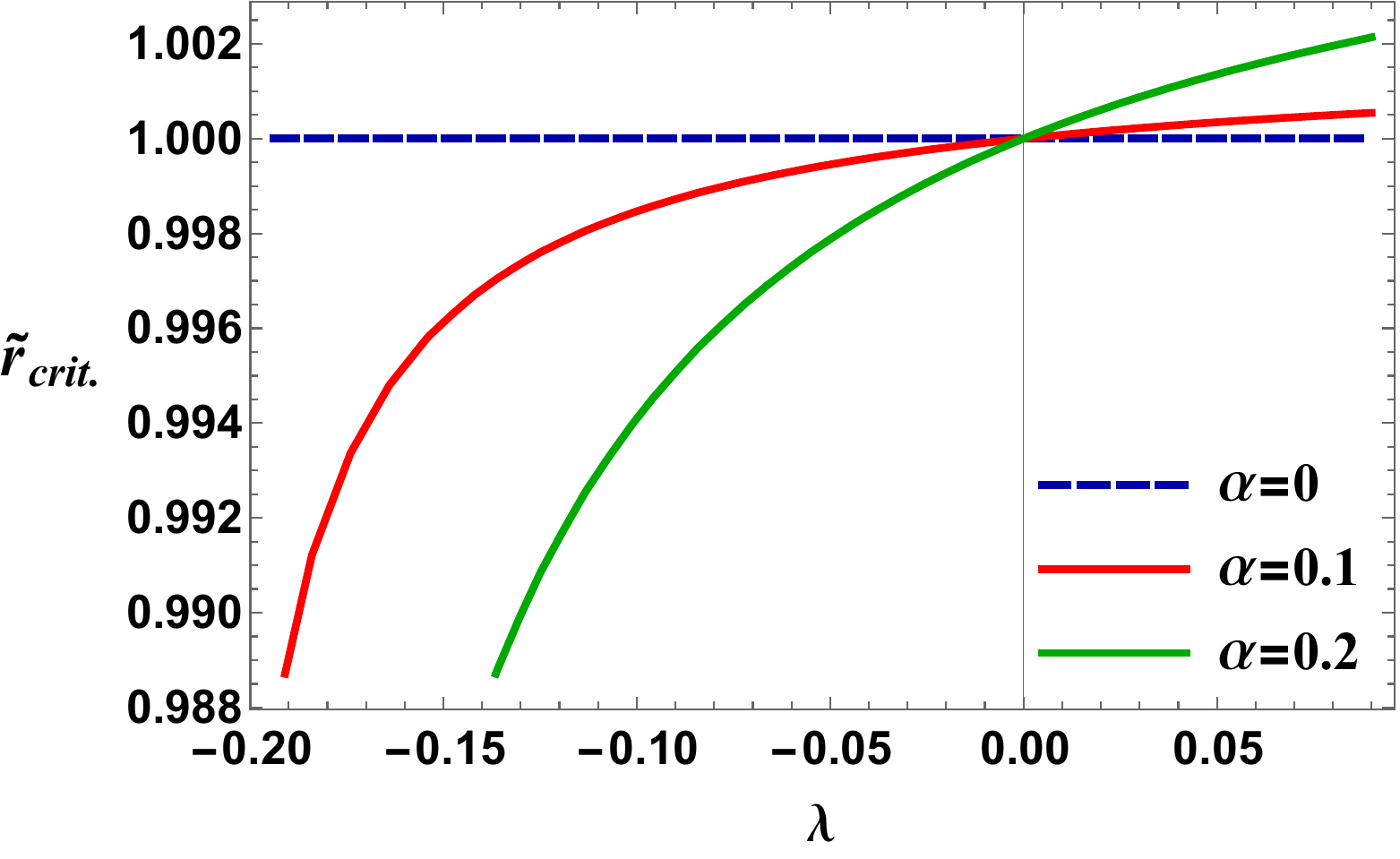}\,\,\,\,\,\,\,\,\,\,\,\,\includegraphics[width=8cm]{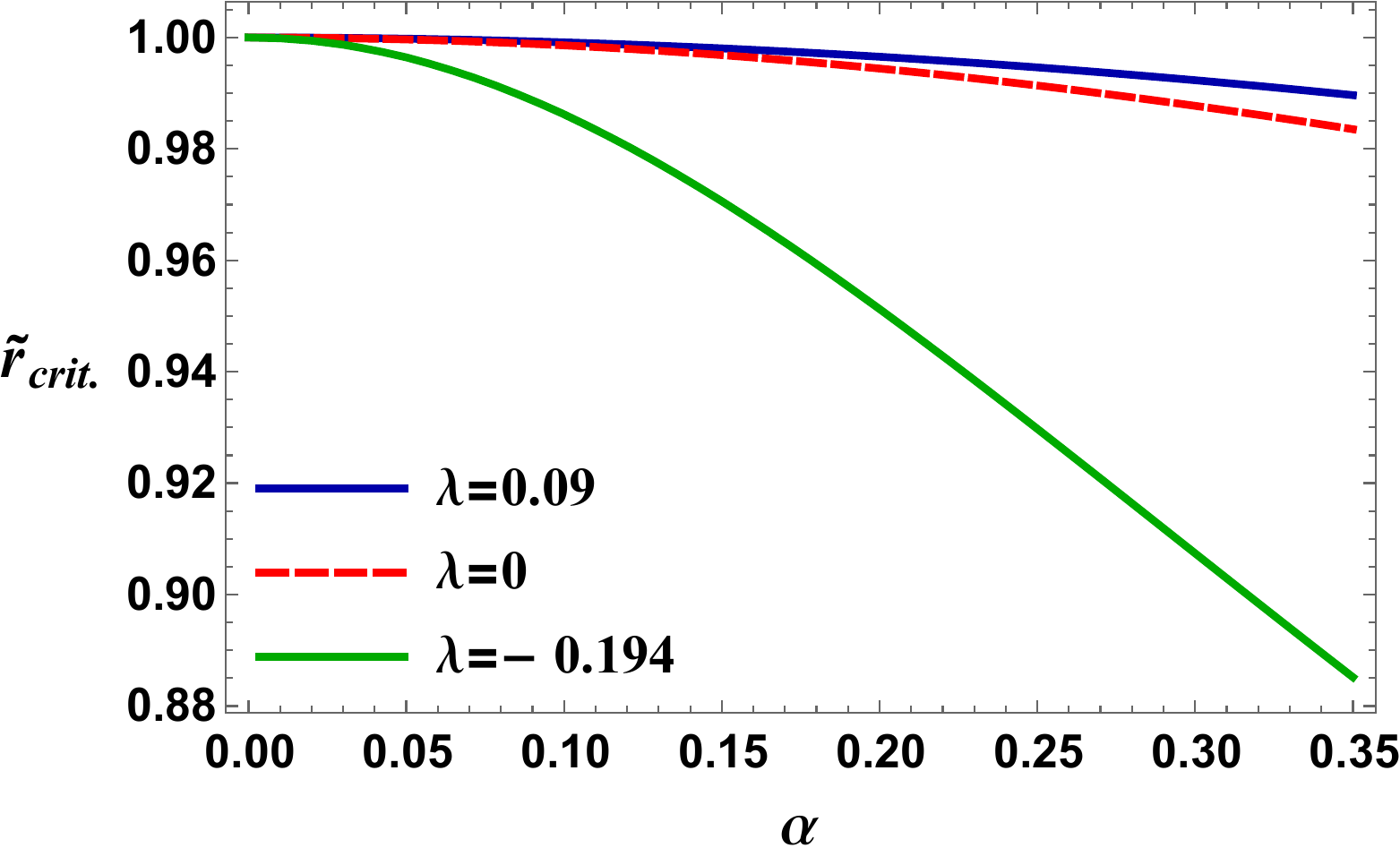}
	\caption{\textit{Left plot}: Normalized transition curve $\tilde{r}_{\mbox{{\tiny crit.}}}=\frac{r_{\mbox{{\tiny crit.}}}}{r_{\mbox{{\tiny crit.}}}^{\lambda=0}}$ as a function of $\lambda$ for fixed values of $\alpha$. \textit{Right plot}: Normalized transition curve $\tilde{r}_{\mbox{{\tiny crit.}}}=\frac{r_{\mbox{{\tiny crit.}}}}{r_{\mbox{{\tiny crit.}}}^{\alpha=0}}$ as a function of $\alpha$ for different values of $\lambda$. It is noted that in both of the figures we set $\ell=1$.  }%\textit{Right plot}: Transition of critical point as a function of $\lambda$ when $\alpha=0.1$. }
	\label{lambda}
\end{figure}

For two separated systems, \emph{e.g.}, $A_1$ and $A_2$, the mutual information would be a proper measure for quantifying the amount of entanglement (or information) that these two systems can share. It is shown that for two separated systems, the mutual information is
a finite quantity and is given by \cite{Casini:2008wt}
\begin{align}\label{HMI}
I\left( {{A_1},{A_2}} \right) = S\left( {{A_1}} \right) + S\left( {{A_2}} \right) - S\left( {{A_1} \cup {A_2}} \right),
\end{align}
where $S\left( {{A_1} \cup {A_2}} \right)$ is the entanglement entropy for the union of two entangling regions. In quantum field theory, mutual information is used as a geometrical regularization of entanglement entropy. It is shown that holographic mutual information undergoes a first order phase transition due to a discontinuity in its first derivative \cite{Headrick:2010zt}. Holographically, this phase transition has in fact a simple geometrical explanation, \emph{e.g.}, for the union of two strips with the same length $\ell$ separated by distance $h$, there are two different configurations which are schematically shown in Fig.\ref{figure1}. It is worth mentioning that we have restricted ourselves to the case in which the entanglement entropy is an increasing function of entangling region. Therefore, the mixed configuration have not been considered \cite{Allais:2011ys}.
 Depending on the value of $h/\ell$ the corresponding minimal configurations may change from one to another and this defines a critical ratio as $r_{crit.}=h/\ell$, in which
	\begin{equation}\label{saa}
	S({A_1} \cup {A_2}) = \left\{ {\begin{array}{*{20}{c}}
		S_{con.} \hfill & {\quad 0 < \frac{h}{\ell} < {r_{crit.}}} \hfill  \\
		S_{dis.} \hfill & {\quad {r_{crit.}} \le \frac{h}{\ell}}. \hfill  \\
		\end{array}} \right.
	\end{equation}
Consequently, the holographic mutual information vanishes or takes a finite value which can be written as
\begin{equation}\label{i2}
I\left( {{A_1},{A_2}} \right) = \left\{ {\begin{array}{*{20}{c}}
	{{2S}\left( \ell \right) - {S}\left( h \right) - {S}\left( {h + 2\ell} \right)}, \hfill & {\quad 0 < \frac{h}{\ell} < {r_{crit.}}} \hfill  \\
	0, \hfill & {\quad {r_{crit.}} \le \frac{h}{\ell}} \hfill  \\
	\end{array}} \right.
\end{equation}
where in AdS$_5$ background one obtains ${r_{crit.}} = \sqrt 3  - 1$. It is noted that in each case $S$ stands for HEE of the corresponding entangling region which is given by \eqref{EE}. Now the aim is to study the effect of  $\alpha$ and $\lambda$ on the mutual information and its phase transition. In the presence of momentum dissipation and GB term, let us write the mutual information as
\begin{equation}
I\left( {{A_1},{A_2}} \right) = {I_0}\left( {{A_1},{A_2}} \right) + \Delta I\left( {{A_1},{A_2}} \right),
\end{equation}
where ${I_0}\left( {{A_1},{A_2}} \right)$ stands for the mutual information when $\alpha=\lambda=0$, and after making use of the corresponding entanglement entropies for $\ell$ , $h$ and $2\ell+h$ regions from \eqref{EE}, it is obtained as
\begin{equation}
{I_0} = \frac{{{H^2}L^3}}{{4{G_N}}}{i_0},\,\,\,\,\,\,\,\,{i_0} = \frac{{4{\pi ^{3/2}}\Gamma {{\left( {\frac{2}{3}} \right)}^3}}}{{\Gamma {{\left( {\frac{1}{6}} \right)}^3}}}\left(\frac{1}{{{{(2\ell  + h)}^2}}} + \frac{1}{{{h^2}}} - \frac{2}{{{\ell ^2}}}\right).
\end{equation}
On the other hand, the correction part becomes
\begin{equation}
\Delta I\left( {{A_1},{A_2}} \right) = \frac{{{H^2}L^3}}{{4{G_N}}}\left( (1 - \frac{3}{2}\lambda ){\alpha ^2}{i_1} + \frac{9}{2}\lambda {i_0} \right) + {\cal O}\left( {{\lambda ^2},{\alpha ^4}} \right),
\end{equation}
where
\begin{equation}
i_1=\frac{1}{4}\log\frac{\ell^2}{h(2\ell+h)}.
\end{equation}
As it is shown in Fig.\ref{mutualfig}, by turning on the momentum relaxation parameter, mutual information between two regions decreases, whereas in a fixed  momentum relaxation parameter, mutual information linearly increases by GB parameter.
\\Moreover, Fig.\ref{lambda} shows the normalized curves of phase transition as a function of GB and momentum relaxation parameters. One observes that for a fixed momentum relaxation parameter, the phase transition of holographic mutual information takes place in larger distance by increasing the GB parameter (left plot in Fig.\ref{lambda}). The general behavior of phase transition is decreasing by $\alpha$, though depending on the sign of GB coupling $\lambda$, it behaves differently; for $\lambda>0$ the phase transition occurs in larger $h$ comparing to the cases of $\lambda\leq 0$.
%  \begin{figure}[h!]
%	\centering
%	\includegraphics[width=8cm]{mutual1.png}%\,\,\,\,\,\,\includegraphics[width=7cm]{onlylambda}
%	\caption{ Normalized transition curve as a function of momentum relaxation parameter $\alpha$ for $\lambda=0.1$ (upper curve), $\lambda=0$ (middle curve) and $\lambda=-0.19$ (lower curve). }%\textit{Right plot}: Transition of critical point as a function of $\lambda$ when $\alpha=0.1$. }
%	\label{AA}
%\end{figure} \\

%But,  there is a discontinuity in the value of $\lambda$, as it is illustrated in above figure (right plot) for specific value of relaxation parameter $\alpha=0.1$, this happens for $\lambda\approx-0.11$. We have plotted  $\tilde{r}_{\mbox{{\tiny crit.}}}$ as a function of $\alpha$ for two values of $\lambda$ in figure \ref{fig3}.

%\begin{figure}[h!]\label{fig3}
%	\centering
%	\includegraphics[width=7cm]{pl-ze-pl.pdf}
%	\caption{ Transition of normalized critical point as a function of relaxation parameter $\alpha$ when $\lambda=0.03$, $\lambda=0$ and $\lambda=-0.3$. }
%\end{figure}

\subsection{Holographic Tripartite Information }

Besides mutual information, in a three-body system with topological order, tripartite information might be utilized as a quantity to characterize entanglement in states of the system. It was first introduced in \cite{Kitaev:2005dm} as the topological entropy and defined by
\begin{align}\label{3par}
{I^{[3]}}\left( {{A_1},{A_2},{A_3}} \right) &= S\left( {{A_1}} \right) + S\left( {{A_2}} \right) + S\left( {{A_3}} \right) - S\left( {{A_1} \cup {A_2}} \right) - S\left( {{A_1} \cup {A_3}} \right) \notag\\
&- S\left( {{A_2} \cup {A_3}} \right) + S\left( {{A_1} \cup {A_2} \cup {A_3}} \right),
\end{align}
where $S\left( {{A_1}\cup{A_2}\cup{A_3}} \right)$ is the entanglement entropy for the union of three subsystems. It is shown that the tripartite information is always finite even when the regions share boundaries. To compute the holographic tripartite information, the union terms of $S(A_i\cup A_j)$ and $S\left( {{A_1}\cup{A_2}\cup{A_3}} \right)$ deserve to be discussed further. For three strips, $A_1$, $A_2$ and $A_3$ of the same length $\ell$ separated by distance $h$, Fig.\ref{imutual1} shows schematically all possible diagrams for computing the union parts of tripartite information.  The rest of the configurations can be obtained by rearranging these ones.
\begin{figure}[h!]
		\centering
	\begin{tikzpicture}[scale=0.55]
    \draw[ultra thick,c2] (3.5,0) -- (4.5,0);
	\draw[ultra thick,c2] (5,0) -- (6,0);
	\draw[ultra thick,c2] (6.5,0) -- (7.5,0);
	\draw[ultra thick,cyan] (4.5,0) arc (0:180:0.5cm);
	\draw[ultra thick,cyan] (6,0) arc (0:180:0.5cm);
    \draw[] (3,0) node[left] {$S_{1}$};	
    \draw[] (4,0) node[below] {$A_1$};
    \draw[] (5.5,0) node[below] {$A_2$};
    \draw[] (7,0) node[below] {$A_3$};

    \draw[ultra thick,c2] (10.5,0) -- (11.5,0);
	\draw[ultra thick,c2] (12,0) -- (13,0);
	\draw[ultra thick,c2] (13.5,0) -- (14.5,0);
	\draw[ultra thick,cyan] (12,0) arc (0:180:0.25cm);
	\draw[ultra thick,cyan] (13,0) arc (0:180:1.25cm);
    \draw[] (10,0) node[left] {$S_{2}$};

    \draw[ultra thick,c2] (17.5,0) -- (18.5,0);
	\draw[ultra thick,c2] (19,0) -- (20,0);
	\draw[ultra thick,c2] (20.5,0) -- (21.5,0);
	\draw[ultra thick,cyan] (20.5,0) arc (0:180:1cm);
	\draw[ultra thick,cyan] (21.5,0) arc (0:180:2cm);
    \draw[] (17,0) node[left] {$S_{3}$};

    \draw[ultra thick,c2] (0,-4) -- (1,-4);
	\draw[ultra thick,c2] (1.5,-4) -- (2.5,-4);
	\draw[ultra thick,c2] (3,-4) -- (4,-4);
	\draw[ultra thick,cyan] (1,-4) arc (0:180:0.5cm);
	\draw[ultra thick,cyan] (2.5,-4) arc (0:180:0.5cm);
    \draw[ultra thick,cyan] (4,-4) arc (0:180:0.5cm);
    \draw[] (-0.5,-4) node[left] {$S_{4}$};	

   \draw[ultra thick,c2] (7,-4) -- (8,-4);
	\draw[ultra thick,c2] (8.5,-4) -- (9.5,-4);
	\draw[ultra thick,c2] (10,-4) -- (11,-4);
	\draw[ultra thick,cyan] (11,-4) arc (0:180:0.5cm);
	\draw[ultra thick,cyan] (8.5,-4) arc (0:180:0.25cm);
    \draw[ultra thick,cyan] (9.5,-4) arc (0:180:1.25cm);
    \draw[] (6.5,-4) node[left] {$S_{5}$};

   \draw[ultra thick,c2] (14,-4) -- (15,-4);
	\draw[ultra thick,c2] (15.5,-4) -- (16.5,-4);
	\draw[ultra thick,c2] (17,-4) -- (18,-4);
	\draw[ultra thick,cyan] (17,-4) arc (0:180:1cm);
	\draw[ultra thick,cyan] (16.5,-4) arc (0:180:0.5cm);
    \draw[ultra thick,cyan] (18,-4) arc (0:180:2cm);
    \draw[] (13.5,-4) node[left] {$S_{6}$};	

    \draw[ultra thick,c2] (21,-4) -- (22,-4);
	\draw[ultra thick,c2] (22.5,-4) -- (23.5,-4);
	\draw[ultra thick,c2] (24,-4) -- (25,-4);
	\draw[ultra thick,cyan] (22.5,-4) arc (0:180:0.25cm);
	\draw[ultra thick,cyan] (24,-4) arc (0:180:0.25cm);
    \draw[ultra thick,cyan] (25,-4) arc (0:180:2cm);
    \draw[] (20.5,-4) node[left] {$S_{7}$};	
    \end{tikzpicture}
		\caption{Schematic representation of competing configurations in the computation of $S(A_i\cup A_j)$ and $S\left( {{A_1}\cup{A_2}\cup{A_3}} \right)$.}
		\label{imutual1}
\end{figure}
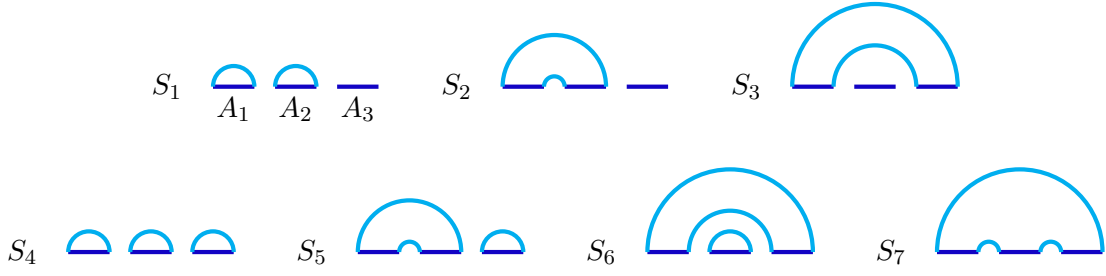 \\
Accordingly one obtains
\[\begin{array}{l}
S\left( {{A_i} \cup {A_j}} \right)\left\{ \begin{array}{l}
2S\left( \ell  \right) \equiv {S_1} \\
S\left( {2\ell  + h} \right) + S\left( h \right) \equiv {S_2} \\
S\left( {3\ell  + 2h} \right) + S\left( {\ell  + 2h} \right) \equiv {S_3} \\
\end{array} \right.
,S\left( {{A_1}\cup{A_2}\cup{A_3}} \right)\left\{ \begin{array}{l}
3S\left( \ell  \right) \equiv {S_4} \\
S\left( {3\ell  + 2h} \right) + S\left( {\ell  + 2h} \right) + S\left( \ell  \right) \equiv {S_5} \\
S\left( {2\ell  + h} \right) + S\left( \ell  \right) + S\left( h \right) \equiv {S_6} \\
S\left( {3\ell  + 2h} \right) + 2S\left( h \right) \equiv {S_7} \\
\end{array} \right. \\
\end{array}\]
Therefore, one can write the following expression for the tripartite information
\begin{equation}\label{tripartite}
{I^{\left[ 3 \right]}}\left( {A_1,A_2,A_3} \right) = 3S\left( \ell  \right) - 2\min \left\{ {{S_1},{S_2}} \right\} - \min \left\{ {{S_1},{S_3}} \right\} + \min \left\{ {{S_4},{S_5},{S_6},{S_7}} \right\}.
\end{equation}
As a special case when $\alpha  = 0$, the \eqref{tripartite} reduces to
\begin{equation}\label{i3}
{I^{\left[ 3 \right]}}\left( {A_1,A_2,A_3} \right) = \left\{ {\begin{array}{*{20}{c}}
	{{S}\left( \ell \right) - 2{S}\left( {h + 2\ell} \right) + {S}\left( {2h + 3\ell} \right)}, \hfill & {\quad 0 < \frac{h}{\ell} < {r_1}} \hfill  \\
	{2{S}\left( h \right) - 3{S}\left( \ell \right) + {S}\left( {2h + 3\ell} \right)}, \hfill & {\quad {r_1} \le \frac{h}{\ell} < {r_2}} \hfill  \\
	0, \hfill & {\quad {r_2} \le \frac{h}{\ell}} \hfill  \\
	\end{array}} \right.
\end{equation}
where in AdS$_5$ background  one finds two critical distances as $r_1=\sqrt 3  - 1$ and $r_2=\frac{{\sqrt 7  - 1}}{2}$, where the value of tripartite information has been changed identically. \\
Similar to the mutual information, one can investigate that in presence of momentum relaxation parameter, the transition curves show a decreasing behavior with respect to $\alpha$ and $\lambda$ and for positive (negative) value of $\lambda$, the phase transition in tripartite information happens in larger (smaller) ratio than the case of $\lambda=0$.

\begin{figure}[h!]
	\centering \includegraphics[width=8cm]{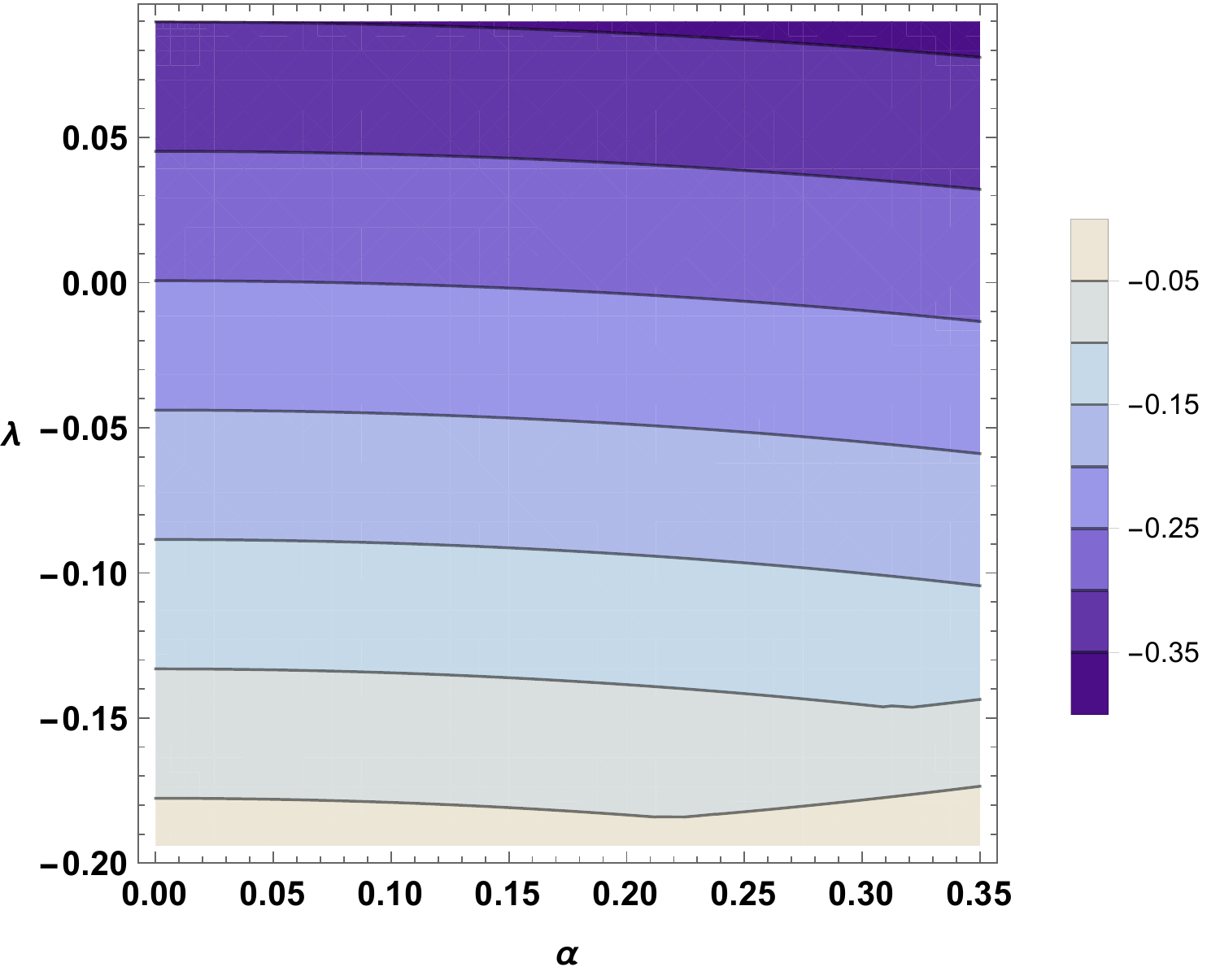}\,\,\,\,\,\,\,\,\,\,\,\,\includegraphics[width=8cm]{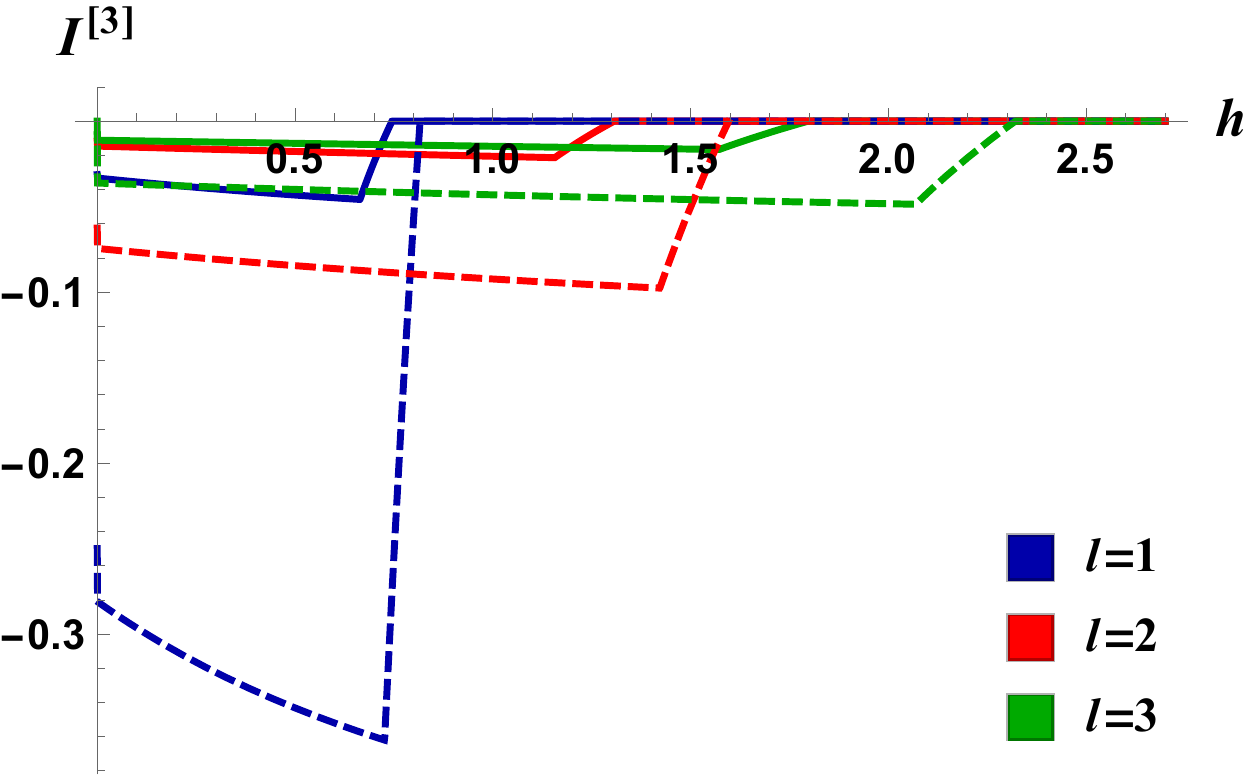}
	\caption{ \textit{Left plot}: Contour plot of tripartite information for $\ell=1$ and $h=0.7$. \textit{Right plot}: Tripartite information for $\alpha=0.3$, $\lambda=-0.194$ (solid curves) and $\lambda=0.09$ (dashed curves). In all ranges that we have considered $I^{[3]}<0$.}\label{fig4}%\textit{Right plot}: Transition of critical point as a function of $\lambda$ when $\alpha=0.1$. }
\end{figure}

We end this subsection with a comment on special property of tripartite information in holographic theories. Tripartite information can be written in terms of the mutual information as follows
\begin{align}\label{3par1}
 {I^{[3]}}\left( {{A_1},{A_2},{A_3}} \right){\rm{ }} = I\left( {{A_1} \cup {A_2}} \right) + I\left( {{A_1} \cup {A_3}} \right) - I({A_1},{A_2} \cup {A_3}).
\end{align}
For arbitrary states of systems, tripartite information has no definite sign, namely depending on the underlying field theory, this quantity can be positive, negative or zero \cite{Casini:2008wt, Hayden:2011ag, Mozaffar:2015xue}. However, in strongly coupled CFTs with holographic duals it is argued that tripartite information is always negative \cite{Alishahiha:2014jxa, Mirabi:2016elb},
and this property is related to the monogamy of the mutual information.\footnote{In the context of quantum information theory, the inequality of the form $F({A_1},{A_2})+F({A_1},{A_3})\leq F({A_1},{A_2}U{A_3})$ is known as monogamy relation. This feature of measurement is related to the security of quantum cryptography indicating that entangled correlations between $A_1$ and $A_2$ cannot be shared with a third system $A_3$ without spoiling the original entanglement \cite{Hayden:2011ag}.} In principle, it can be concluded that the holography leads to a constraint on this quantity and its sign might be employed in various works (see for example \cite{Pastawski:2015qua, Almheiri:2014lwa}). In Fig.\ref{fig4}, we have plotted the tripartite information as a function of momentum relaxation parameter and GB coupling. One observes that it always remains negative. This behavior also holds when one changes the length of entangling regions for the given (fixed) values of momentum relaxation and GB coupling parameters.

%%%%%%%%%%%%%%%%%%%%%%%%%%%%%%%%%%%%%%%%%%%%%%%%%%%%%%%%%%%%%%%%%%%%%%%%%
\subsection{Wilson Loop}

The Wilson loop has some similar properties as the entanglement entropy and it can be used to  investigate  the phase transitions in quantum systems. Namely this quantity characterizes phases of gauge theories in terms of the potential between electric charges. The expectation value of the Wilson loop which is related to the effective potential between a quark and antiquark pair, can be approximated by the gauge/gravity correspondence as \cite{Maldacena:1998im}
\begin{equation}
\langle W\rangle\sim e^{-\frac{A(\Sigma)}{2\pi\alpha'}},
\end{equation}
where $(2\pi\alpha')^{-1}$ is the string tension, $\Sigma$ is the string world sheet that extends in the bulk and $A(\Sigma)$ stands for the Nambu-Goto action for the string which by saddle point approximation reduces to the minimal surface of the classical string. Thus for a rectangular Wilson loop of width $\ell$, the corresponding potential between the quark and antiquark is given by
\begin{equation}
V=\frac{L^2}{\pi\alpha'}\int_\epsilon^{\rho_t} d\rho\frac{\rho_t^2}{\rho^2}\sqrt{\frac{f(\rho)}{f(\rho)\rho_t^4-\rho^4}},
\end{equation}
where $f(\rho)$ is defined by \eqref{frho}. After doing the same computation as previous section, we find that
\begin{equation}
V = V_0^{AdS}+\Delta V=\frac{L^2}{\alpha'}\left[ \frac{1}{\epsilon }- \frac{2 \Gamma \left(\frac{3}{4} \right)^2}{\Gamma \left( \frac{1}{4} \right)^2}\frac{1}{\ell }\right] +\frac{L^2}{\alpha'}\left[\frac{  \Gamma\left( \frac{3}{4} \right)^2}{\Gamma\left( \frac{1}{4} \right)^2}\frac{ \lambda}{\ell } - \frac{{\pi\left( {3\lambda  + 2} \right) {\alpha ^2}\ell }}{{192 \,\Gamma {{\left( {\frac{3}{4}} \right)}^4}}} \right].
\end{equation}
Therefore, after regularization and subtracting the effective potential of the AdS vacuum, the first-order correction of force between the quark and antiquark in this model is found as follows
\begin{equation}
\Delta F=-\frac{d}{d\ell}\Delta V=\frac{L^2}{\alpha'}\left[\frac{  \Gamma\left( \frac{3}{4} \right)^2}{\Gamma\left( \frac{1}{4} \right)^2}\frac{\lambda}{\ell^2 } + \frac{{\pi\left( {3\lambda  + 2} \right) {\alpha ^2} }}{{192 \,\Gamma {{\left( {\frac{3}{4}} \right)}^4}}} \right].
\end{equation}
Noting that from the regularized part of AdS one receives an attractive force between these external particles, the correction due to the momentum relaxation is always repulsive and independent of $\ell$ which is in agreement with the results in \cite{Mozaffara:2016iwm}. However, the contribution of GB coupling is somehow nontrivial. The $\lambda$-correction part depends on separation $\ell$ and according to the sign of GB coupling could be either positive or negative which results in decreasing or increasing attractive force between quark and antiquark, respectively.   % We have depicted in Fig.\ref{loop}, the correction part of force with respect to $\ell$ for a fixed momentum relaxation parameter which indicates that for $\lambda>0$ it is repulsive while for $\lambda<0$ it is attractive.
%\begin{figure}[h!]
%	\centering \includegraphics[width=9cm]{wilson}%\,\,\,\,\includegraphics[width=3cm]{adad.png}
%	\caption{The correction part of force ($\Delta F$) as a function of $\ell$ with $\alpha=0.1$ for positive and negative values of $\lambda$.  }\label{loop}
%\end{figure}

%********************************************************************************************************

\section{Conclusion}

In principle, in the holographic models, considering higher-curvature terms in the gravity action is well motivated for reasons; in particular, addressing different types of central charges could be an example. The Lovelock gravity is indeed the simplest set of higher derivative terms
in which various Euler densities appear as higher derivative interactions in the gravity theory.

In this paper, we studied the effect of higher-order derivative terms on some nonlocal probes in the theories with momentum relaxation parameter. There are in fact two kinds of deformation in the states of dual field theory in this model: the higher-curvature terms, which could address the low-energy quantum excitation corrections, and the deformation due to scalar fields, which are responsible for the momentum conservation breaking. We used holographic methods to obtain the corresponding changes due to these deformations in the coefficient of universal part in entanglement entropy. Higher-order gravity theories are interesting in a sense that they provide us with an effective description of quantum corrections and one may probe the finite coupling effects and the $a$- and $c$-theorems via making such corrections to the Einstein gravity theory in the bulk space. We used five-dimensional Einstein GB gravity together with three spatial-dependent massless scalar fields to obtain the corrections to universal and finite parts of HEE for strip, spherical and cylindrical entangling regions. For an interval of length $\ell$ on an infinite line, Myers and Singh introduced a candidate for $c$-function in a $d$-dimensional CFT which is the coefficient of the finite term  in entanglement entropy. This expression in $d=4$ is given by \eqref{cfunction} and it can be considered as a function of the anomaly coefficients in the underlying CFT. We showed that in the presence of momentum relaxation parameter and GB coupling, this expression has been modified as \eqref{myers}. Moreover, in computing the HEE for a strip, a universal logarithmic term appears due to the momentum relaxation parameter which has been modified by the GB coupling. This universal term  vanishes at $\lambda\backsimeq0.66$; however, noting that the GB coupling is constrained to a small range, \emph{i.e.} $-0.194\lesssim\lambda\leq 0.09$ \cite{Buchel:2009sk, Hofman:2008ar},  one gets a positive valued universal term due to both the momentum relaxation and GB term in the present range\footnote{It is worth mentioning that considering the GB terms non-perturbatively leads to the violation of causality in any pure Gauss-Bonnet gravity \cite{Camanho:2014apa}. Moreover, we assume that momentum relaxation does not change the constraints on the GB coupling. We thank the referee for his/her useful comment on this point.}.  \\
In the case of spherical entangling region, the coefficient of universal term in HEE could potentially address the $a$-central charge of corresponding dual conformal field theory whereas the $c$-central charge is related to the coefficient of universal term in HEE for cylindrical entangling region. For theories dual to Einstein gravity one obtains $a=c$; however, in the case of GB gravity one obtains unequal $a$ and $c$, this is indeed the main motivation of considering such term in the gravity action. We obtained the modified coefficients of universal terms which can be interpreted as `$c$'-type central charge of dual field theory.\\
In the context of quantum information theory and also quantum many-body systems, for two disjointed systems, the mutual information is usually used as a measure of quantum entanglement that these two systems can share; the mutual information can also be utilized as a useful probe to address certain phase transitions and critical behavior in these theories. For example, it is known that mutual information undergoes a transition beyond which it is identically zero; this kind of transition, which is called as disentangling transition, is in fact universal qualitative feature for all classes of theories with holographic duals \cite{Fischler:2012uv}. In this paper, we considered the effect of GB term on such phase transition in both of the mutual and tripartite information and it was shown that the behavior of such phase transition is different depending on the sign of GB coupling. For two strips with same length separated by distance $h$, we showed that in a fixed momentum relaxation parameter, the phase transition of holographic mutual information takes place in larger distance by increasing the GB parameter. The general behavior of phase transition is decreasing by $\alpha$, though for $\lambda>0$ the phase transition occurs in larger $h$ comparing to the cases of $\lambda\leq 0$. For $\lambda>0$ this transition happens in larger value than the case of $\lambda<0.$ We also showed that the tripartite information has negative value in our setup which means that mutual information is monogamous. 

Moreover, by considering the holographic Wilson loop, we found that the sign of $\lambda$ plays a crucial role in computation of the effective potential and its corresponding force between point-like external objects. The result shows that both momentum
dissipation and GB coupling parameters can lead to correction of the potential and corresponding force  between quark and anti quark. Noting that from the regularized part of AdS one receives an attractive force between these external particles, the correction due to the momentum relaxation is always repulsive and independent of $\ell$ which is in agreement with the results in \cite{Mozaffara:2016iwm}. However, the contribution of GB coupling is somehow nontrivial. The $\lambda$-correction part depends on separation $\ell$ and according to the sign of GB coupling could be either positive or negative which results in decreasing or increasing attractive force between quark and antiquark, respectively.

%********************************************************************************************************
\section*{Acknowledgments}

The author would like to thank Mohsen Alishahiha, M. Reza Mohammadi-Mozaffar and Ali Mollabashi for their helpful comments and discussions. MRT also wishes to acknowledge A. Akhvan, A. Faraji, A. Naseh, A. Shirzad, F. Omidi, F. Taghavi and M. Vahidinia for some their comments. We also thank the referees of this paper for their useful comments. This work has been supported in parts by IAUCTB.

\section*{Appendix}
In this appendix, we write down some related computation of finding the entropy functional \eqref{efunc}. In the present case there are two orthogonal normal vectors as follows
\begin{equation}\label{unitnormals}
\begin{array}{l}
\Sigma_1\,\,\,:\,\,\,\,t=0\,\,\,\,\,\,\,\,\,\,\,\,\,\,\,\,\,\,\,\,\,\,\,\,\,\,\,\,\,\,\,\,
\,\,\,\,{n_1} =  \left\{ {\frac{{\sqrt f {L}}}{\rho },0,0,0,0} \right\}, \\
\Sigma_2\,\,:\,\,\,x_1-x(\rho)=0\,\,\,\,\,\,\,\,\,\,\,\,\,\,\,\,{n_2} =  \left\{ {0, - \frac{{x'{L}}}{{\rho \sqrt {f{{x'}^2} + 1} }},\frac{{{L}}}{{\rho \sqrt {f{{x'}^2} + 1} }},0,0} \right\}. \\
\end{array}
\end{equation}
The corresponding extrinsic curvatures of the hypersurface are given by
\begin{equation}
\mathcal{K}_{\mu \nu }^{(1)} = 0,\,\,\,\,\,\,\,\,\,\,\,\,\,\,\,\,\mathcal{K}_{\mu \nu }^{(2)} = {L}\left( {\begin{array}{*{20}{c}}
	0 & 0 & 0 & 0 & 0  \\
	0 & {{C_1}{f^{ - 1}}} & {{C_1}x'} & 0 & 0  \\
	0 & {{C_1}x'} & {{C_1}f{{x'}^2}} & 0 & 0  \\
	0 & 0 & 0 & {{C_2}} & 0  \\
	0 & 0 & 0 & 0 & {{C_2}}  \\
	\end{array}} \right),
\end{equation}
where
\begin{equation}
{C_1} = \frac{{2\left( {1 + fx{'^2}} \right)fx' - \rho \left( {f'x' + 2fx''} \right)}}{{2{\rho ^2}{{(1 + fx{'^2})}^{5/2}}}},\,\,\,\,\,\,\,\,\,\,\,\,\,\,\,\,{C_2} = \frac{{fx'}}{{{\rho ^2}\sqrt {1 + fx{'^2}} }}.
\end{equation}
Consequently, for a strip entangling region, the entanglement entropy of \eqref{fursaev} for a general five-dimensional higher-curvature gravity theory becomes
\begin{equation}\label{eefunctional}
S = \frac{{{H^2}L^3}}{{4{G_N}}}\int d \rho \frac{{\sqrt {{{x'}^2} + {f^{ - 1}}} }}{{{\rho ^3}}}\left( {1 + \mathcal{A} + \mathcal{B}} \right),
\end{equation}
where
\begin{equation}
\begin{array}{l}
\mathcal{A} = \frac{{ - 8\left( {10a + 2b + c} \right)f + \left( {32a + 7b + 4c} \right)\rho f' - \left( {4a + b} \right){\rho ^2}f''}}{{2L^2}} + \frac{{\left[ {\left( {3b + 4c} \right)\rho f' - \left( {b + 4c} \right){\rho ^2}f''} \right]fx{'^2}}}{{2L^2\left( {1 + fx{'^2}} \right)}}, \\
\\
\mathcal{B} =  - \frac{{{\rho ^4}\left[ {b{{\left( {2{C_2} + {C_1}\left( {1 + fx{'^2}} \right)} \right)}^2} + 4c\left( {2{C_2}^2 + {C_1}^2{{\left( {1 + fx{'^2}} \right)}^2}} \right)} \right]}}{{2L^2}}. \\
\end{array}
\end{equation}
By fixing the coupling constants of higher-order terms in \eqref{eefunctional} according to five-dimensional GB gravity, one obtains the entropy functional \eqref{efunc}.

\end{document}